\documentclass[11pt,twoside]{article}
\usepackage{newpasp}
\usepackage{graphicx}
\usepackage{epsf}

\def\nii{NII}
\def\hei{HeI}
\def\heii{HeII}
\def\oi{OI}
\def\oii{OII}
\def\oiii{OIII}
\def\sii{SII}

\def\asp{ASP}

\index{Bland-Hawthorn, J.}
\index{Putman, M. E.}

\begin{document}

\title{The Galactic Halo UV Field, Magellanic Stream and HVCs}
\author{J. Bland-Hawthorn}
\affil{Anglo-Australian Observatory, PO Box 296, Epping, NSW 1710, Australia}
\author{M.E. Putman}
\affil{Australia Telescope National Facility, PO Box 76, Epping, NSW 1710, Australia}

\begin{abstract}
Significant numbers of high-velocity \HI\ clouds (HVCs) have now been
detected in H$\alpha$, with a subset seen in low ionization lines
(e.g. [\nii]).  It was originally hoped that the observed H$\alpha$ strength
would provide a distance constraint to individual clouds. This idea
requires that a useful fraction ($f_{\rm esc} > 1$\%) of ionizing
photons escape the Galaxy, and that the halo ionizing field is
relatively smooth, as we discuss. HVCs which are known to be close to
the Sun are H$\alpha$-bright; the brightest clouds also show enhanced [\nii]
emission, in contrast to the Magellanic Stream where the low
ionization emission lines are always weak compared to H$\alpha$. But an
acute complication for H$\alpha$ distances is the apparent H$\alpha$ brightness
of the Magellanic Stream along several sight lines, comparable or
brighter than local HVCs. To account for this, we present three
possible configurations for the Magellanic Stream and propose a
follow-up experiment.  If we normalize the distances to local HVCs,
some HVCs appear to be scattered throughout the Galactic halo on
scales of tens of kiloparsecs.
\end{abstract}

\vspace{-20pt}
\section{The big picture}
\vspace{-5pt}

At this meeting, it was clear that the intergalactic medium has become
a key frontier across a range of disciplines within astrophysics. The
equation of state of the IGM is influenced by the spectral energy
distribution of the diffuse cosmic UV field. It is therefore
unfortunate that the far UV is the most uncertain part of the cosmic
spectrum because it is very difficult to measure at any redshift
(Henry 1991). It is normally inferred from the `proximity effect' in
quasar spectra, or from reprocessed recombination flux, but these
bounds are highly uncertain (Kulkarni \& Fall 1993). At low redshift,
an alternative scheme is to add up the UV escape fraction ($f_{\rm
esc}$) from likely sources, but again the estimates are highly
controversial. If $f_{\rm esc}$ in star-forming galaxies exceeds a
few percent, then massive stars probably dominate over black-hole
processes in producing the ionizing background (Giallongo, Fontana \&
Madau 1997).

In this invited review, we present evidence for UV escape from
galaxies (\S 2,3,4).  This conference has devoted a special session to
high-velocity clouds, and so we use the second half of the paper to
discuss the role of H$\alpha$ distances. If the Galactic halo UV field is
sufficiently strong, the observed H$\alpha$ flux from \HI\ clouds can
provide crude distance constraints for each cloud (\S 5).
Bland-Hawthorn \& Maloney (1999$a$, hereafter B99$a$) attempted to
derive $f_{\rm esc}$ indirectly from H$\alpha$ along the Magellanic Stream,
but it now seems that UV from massive stars does not dominate at the
distance of the Stream. Local HVCs with established distance bounds
may provide a better calibration of $f_{\rm esc}$, but significant
uncertainties remain as we show. H$\alpha$ distances may already reveal
that some HVCs are dispersed throughout the Galactic halo (\S 6).  In
\S 7, we discuss possible explanations for the Stream emission and
propose observational tests.

\vspace{-10pt}
\section{Do ionizing photons escape star forming regions?}
\vspace{-5pt}

This topic remains highly controversial.  About the only issue not in
dispute is that something is ionizing and heating the gas in the
haloes of star-forming galaxies. Enhanced temperatures and high levels
of ionization are frequently observed in filaments and in diffuse line
emission several kiloparsecs from the galactic plane (Rand 2000). The
radiation fields from the cosmic background, galaxy group or the hot
galactic halo are much too weak (Maloney 1993; Maloney \&
Bland-Hawthorn 1999, hereafter MB).

We know that a sizeable fraction of UV photons do {\it not} escape
star forming regions. Bronfman et al.\ (2000), after selecting
embedded massive stars from their distinctive FIR colours, show that
about 5.6$\times 10^{52}$ ionizing photons per second are required to
generate the total FIR luminosity. This is about $20-25$\% of the
expected photon rate from the Galaxy (B99$a$; McKee \& Williams
1997). Of the remaining photons, a large fraction is presumably
absorbed within more diaphanous star-forming complexes or within the
diffuse medium.

The Reynolds layer in the Galaxy, and its counterpart in external
galaxies, appears to require at least 15\% of the O (and B\footnote{O
star UV luminosities are correct to within 50\%, but B stars are much
more uncertain and could contribute a useful fraction of ionizing
photons at high latitudes.}) star ionizing flux, and close to 100\% of
the supernova kinetic energy (Reynolds 1984; Domg\"{o}rgen \& Mathis
1994; Hoopes \& Walterbos 2000). Since we know that UV does manage to
escape some \HII\ regions (Rubin et al.\ 1991; Oey \& Kennicutt 1997),
the jury has tended to side with hot, young stars while accepting that
it is unclear how the UV manages to get out.

An argument used against UV escaping the Galaxy altogether (M. Walker,
private communication) is that the \HI\ halo gas extends vertically to
4~kpc in NGC 891 (Swaters et al.\ 1997). But the diffuse plasma shows
high ionization out to 5 kpc (Rand 1997).  Swaters' \HI\ data is at
relatively low resolution (15\arcsec) and is integrated through the
galaxy over a very long baseline ($\sim$20~kpc). The high resolution
dust pictures of Howk \& Savage (2000) reveal that the halo gas is
very filamentary. For the Galaxy, Koo, Heiles \& Reach (1992) have
found related \HI\ structures extending from the plane.

Some of the best evidence that UV must escape the Galaxy comes from
the measured electron density profile from halo pulsars. Manchester \&
Taylor (2000; see also Nordgren, Cordes \& Terzian 1992) have modelled
this with a scale height of 800~pc which exceeds or is comparable to
the scale height of the diffuse \HI\ (warm neutral medium; Lockman
1984). Without fine-tuning, it is unlikely that the Reynolds layer
represents a radiation-bounded medium within a co-extensive \HI\
envelope. We know that the radiation field must be soft from the
weakness of \hei$\lambda$5876 and non-detection of \heii$\lambda$4686
(Reynolds \& Tufte 1995). Furthermore, the observed weakness of
[\oi]$\lambda$6300/H$\alpha$ indicates two things: (i) the ionization
fraction must be high (Reynolds 1989), (ii) all of the UV photons
produced in the disk cannot be absorbed in radiation-bounded \HII\
regions (Domg\"{o}rgen \& Mathis 1994).

The escape problem may be overstated in the context of the most
powerful star-forming complexes which are presumably responsible for
most of the UV production. We now know that the spirals with extended
H$^+$ haloes are those with high star formation rates (Rand 1996;
Lehnert \& Heckman 1996).  It may be that the processes which propel
gas into the halo are the same processes which help UV get out.  The
observed filaments along the minor axis of superwind galaxies show a
clear signature of OB star photoionization and gradual dilution with
altitude ({\it q.v.} Greve et al.\ 2000).  On smaller scales, this may
be what is happening around individual star-forming complexes which
are expected to produce a complex network of superbubbles or chimneys
bursting out of the stratified medium (Rosen, Bregman \& Kelson 1996;
Mac Low 1998; Shelton 1998). Indeed, Veilleux (2000, this meeting)
presented a spectacular HST H$\alpha$ image of NGC 3079 which supports
this: hundreds of vertical filaments are seen emanating from
star-forming complexes across the entire optical disk.

\vspace{-10pt}
\section{Estimates of $f_{\rm esc}$}
\vspace{-5pt}

From observations of four UV-bright starburst galaxies with the {\it
Hopkins Ultraviolet Telescope}, Leitherer et al. (1995) determined
upper limits to $f_{\rm esc}$ of 1, 2, 5 and 15\%, (for Mrk 496, IRAS
08339+6517, Mrk 1267 and Mrk 66, respectively) and concluded that the
escaping fraction must be small. Their analysis did not take into
account the absorption of ionizing radiation by the Galaxy, however,
and as shown by Hurwitz, Jelinsky \& Dixon (1997), this correction
raises the above limits to 3, 5, 11, and 57\%, so that they no longer
provide reliable constraints. (Note also that Hurwitz et al.\ did not
include the effects of absorption by molecular hydrogen, which could
raise these upper limits further; see the discussion in \S 4 of their
paper.) More recently, Steidel, Pettini \& Adelberger (2000) have
determined that a significant fraction of ionizing photons ($f_{\rm
esc} \ga$ 7\%) escape Lyman break galaxies at $z\sim 3.4$.

\begin{figure}
\centerline{ \rotatebox{-90}{\includegraphics[width=15cm]{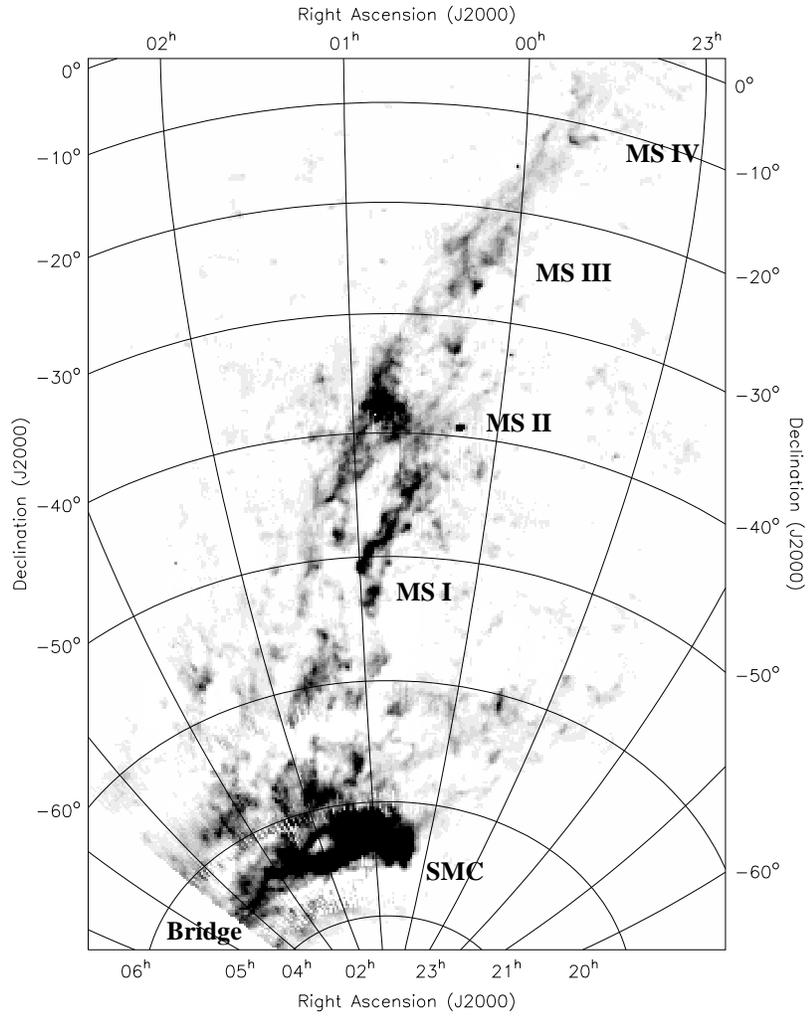} } }
\caption{\small The LMC, SMC, Magellanic Stream and part of the Leading Arm
from the HIPASS \HI\ survey (Putman et al.\ 2000).}
\label{bh2_fig1}
\vspace{-5pt}
\end{figure}

B99$a$ pinned their hopes on the Magellanic Stream
(Fig.~\ref{bh2_fig1}), which passes over the South Galactic Pole, for
an improved estimate of $\hat{f}_{\rm esc}$ ($\approx 6$\%) but this
model underestimates the required flux by at least a factor of
5. (Their escape value $\hat{f}_{\rm esc}$ is defined orthogonally to
the disk plane; the solid-angle averaged value is $f_{\rm esc}$
$\approx$ 1$-$2\%.) It now seems that the Stream H$\alpha$ cannot
arise from the disk UV field (although see \S 7) not least because
Weiner et al.\ (2000) find that certain Stream pointings are much
brighter than the values modelled by B99$a$ (see also Putman et al.\
2000). However, it turns out that the originally derived $\hat{f}_{\rm
esc}$ is within the required range to explain nearby HVCs (\S 6),
assuming they are photoionized by disk stars.

Theoretical models of the transport of ionizing radiation within an
idealized Galactic disk suggest that approximately 10\% of the
ionizing photons produced within the Galaxy escape the disk entirely
(Dove, Shull \& Ferrara 2000; Fransson \& Chevalier 1985; Miller \&
Cox 1993; Bregman \& Harrington 1986; Domg\"{o}rgen \& Mathis 1994).
Another theoretical approach which should be considered is a fractal
gas distribution superimposed on the McKee-Ostriker 3-phase model of
the ISM.

\vspace{-10pt}
\section{The halo UV field: smooth or patchy?}
\vspace{-5pt}

\begin{figure}
\plottwo{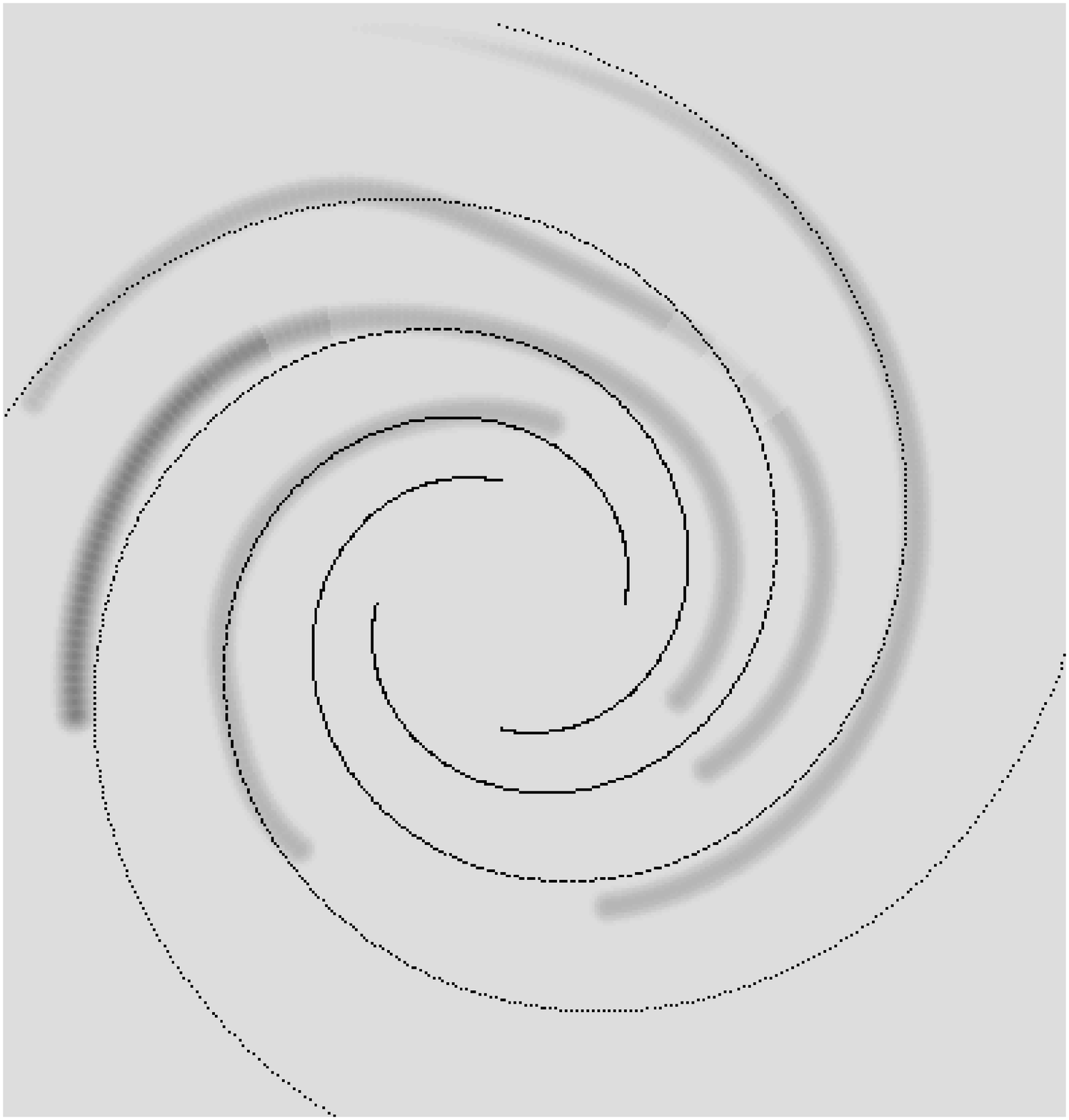}{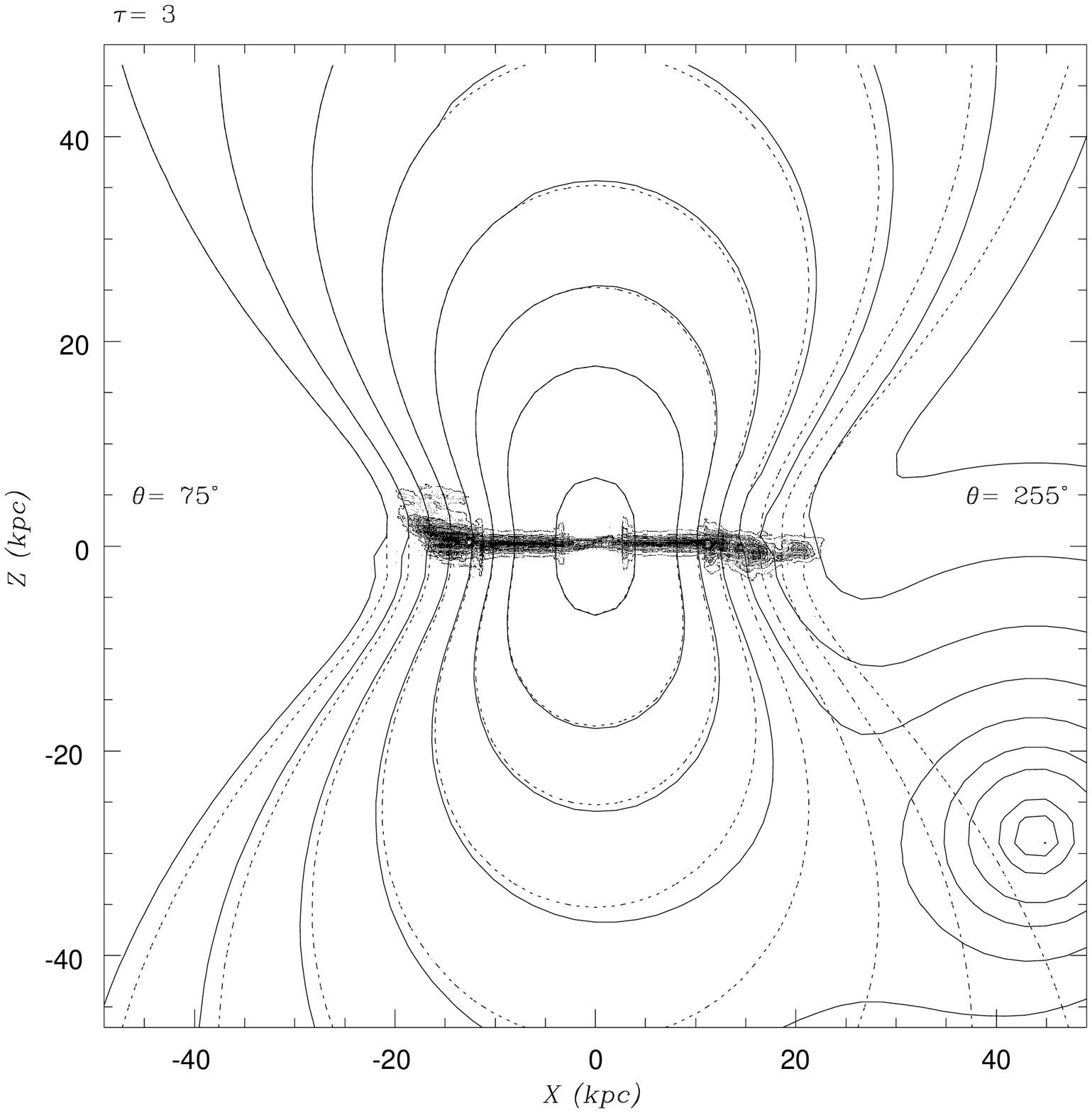}
\caption{\small (a) The locations of the Galaxy's spiral 
arms (Taylor \& Cordes 1993) in relation to the Sun's position 
($\odot$). (b) The Galactic halo ionizing field in a plane 
perpendicular to the Galactic plane (see B99$b$ for details): 
the contours, from outside in, are
approximately $\log\varphi$ $=$ 4, 4.25, 4.5, 4.75, 5, 5.25, 5.5, 6 where 
$\varphi$ is the ionizing photon flux in units of cm$^{-2}$ s$^{-1}$.  }
\label{bh2_fig2}
\vspace{-10pt}
\end{figure}

The structure of the UV halo field is particularly important and
should help to resolve several key issues.  A smooth ionizing halo is
more likely to provide a useful distance constraint through the
surface H$\alpha$ emission.  In Fig.\ref{bh2_fig2}$b$, we show the
predicted halo field for a smooth exponential disk with uniform dust
opacity and an escape fraction of $\hat{f}_{\rm esc} \,=$ 6\% normal
to the disk. (The solid-angle averaged escape fraction in this model
is $f_{\rm esc}\,\approx \,1-2$\%.) If the halo field is very patchy,
then the H$\alpha$ distance constraint may be limited to arguing that
some objects are within the sphere of influence of the Galaxy
(detections) and some are further afield (non-detections).

In external galaxy haloes, Rand (1999; 2000) generally sees smooth
rising trends in [\sii]/H$\alpha$ and [\nii]/H$\alpha$ with $\vert z
\vert$, suggesting a smooth, global ionizing source. But local
variations are also seen when comparing parallel slits, and along a
given slit on either side of the galaxy plane.  Collins \& Rand (2001)
note that the local variations may be specific to filaments (rather
than the diffuse emission) and seem to require an additional source of
ionization.

The UV field from massive stars alone is expected to be patchy. In her
PhD thesis, Cianci (2001) has compared optical line maps with UV
images of spirals observed by the Ultraviolet Imaging Telescope (UIT)
aboard the Space Shuttle. In all cases, there is a 90\% match or better between
UV-identified and H$\alpha$ -identified \HII\ regions. In other words, the
UV images are almost indistinguishable from the H$\alpha$ maps which often
means that clear spiral arms are seen, particularly in late-type
spirals.  For the Galaxy, the locations of the spiral arms ---
particularly the direction of the tangent points --- are fairly well
defined (see Fig.~\ref{bh2_fig2}$a$; Taylor \& Cordes 1993) and this
has important consequences for H$\alpha$ distance constraints (\S 6).

The most energetic \HII\ regions in spirals generally appear to be
`naked', i.e.\ dust does not appear to disrupt their
morphology. Indeed, in his comments after the talk, R.~Allen noted
that this has been known a long time. In a series of papers, Rozas,
Knapen and Beckman (e.g.  Beckman et al.\ 1999) argue that the \HII\
region H$\alpha$ luminosity function appears to have a natural break at
high luminosity. Their preferred explanation is that UV escapes the
most energetic systems.

The UIT images all show evidence of a diffuse inter-arm UV component.
But the UIT bands are set at lower energies than Lyman continuum
photons and therefore probably include a major contribution from AF
stars. In most cases, the UIT images do not rule out the possibility
of a more dispersed, low surface brightness disk component, either
from runaway OB stars (Lynds 1980; Hoogerwerf et al.\ 2000) or field O
stars (Patel \& Wilson 1995; {\it q.v.} Massey et al.\ 1995).

Notably, Slavin, McKee \& Hollenbach (2000) have recently proposed
that cooling hot gas in old supernova remnants could contribute
significantly to the diffuse ionizing field in the disk. This may
appear to contradict our earlier statement (\S 2) that halo coronal
gas provides only a weak EUV field. But the coronal halo parameters
(B99$a$) produce a field that is harder than Slavin's field and is
therefore much less efficient in ionizing hydrogen. Slavin et al.\
find that roughly a third of the original supernova explosion energy
can re-emerge at the remnant stage as a diffuse EUV field.  But we
note that this phase may in fact help the O star flux reach the outer
halo, and may even help to smooth out the disk UV field.

\vspace{-10pt}
\section{A circular problem}
\vspace{-5pt}

The H$\alpha$ distance constraint (Bland-Hawthorn et al.\ 1998, hereafter
B98) was originally formulated to give some indication of whether HVCs
are a relatively local phenomenon or distributed on scales of tens of
kiloparsecs. But it relies on our knowing the mean intensity and
distribution of the halo ionizing field, which in turn relies on H$\alpha$
from an \HI\ screen of known distance, covering fraction, topology and
orientation to our line of sight (Bland-Hawthorn \& Maloney 1999$b$,
hereafter B99$b$). We originally considered the Magellanic Stream
which goes directly over the South Galactic Pole and has been detected
in H$\alpha$ (Weiner \& Williams 1996), but certain parts of the Stream now
appear to be too bright to be explained by the disk UV field (\S 7).

D.W. Sciama (1997, personal communication; see also Bregman 1999) proposed
using high-velocity clouds to estimate $f_{\rm esc}$, but at that time, 
there were few with useful distance constraints. The 
situation really has not improved much over the last three years. 
From the clouds which have been detected,
we derive $\hat{f}_{\rm esc} \,\approx \, 3-12$\% (solid-angle
averaged, $f_{\rm esc}\, \approx\, 1-4$\%) by comparing these data to the
predicted emission measures from a smooth exponential disk model
and a spiral arm model (Bland-Hawthorn \& Maloney 2001, in prep.).

There are several concerns with our application to HVCs. Note that most
clouds are located within 10~kpc of the Sun's position. For a realistic
model, we would at least expect that HVCs `outside' are generally fainter 
than HVCs `within' the Solar Circle, with or without the presence of
spiral arms (see Fig.~\ref{bh2_fig2}$a$). The influence of the Solar Circle 
is barely evident from so few clouds.

Another issue is that at least two HVCs (B98; Putman et al.\ 2000)
have elevated [\nii]/H$\alpha$ emission. Here, we are forced to assume
that the [\nii] emission indicates simply an enhanced electron
temperature (Reynolds, Haffner \& Tufte 1999), rather than the
presence of a more pernicious source (e.g.\ shock heating). There is a
variety of ways to produce this effect, e.g. photoelectric heating
(Wolfire et al.\ 1995).  This may not be a major concern, since after
all we know that high latitude gas in spirals shows enhanced
low-ionization emission (Haffner, Reynolds \& Tufte 1999; Veilleux et
al.\ 1995). In essence, we can use the elevated [\nii]/H$\alpha$ to
argue that some HVCs are more than several kiloparsecs from the plane,
and comprise part of the extended ionized atmosphere seen in external
galaxies.

\vspace{-10pt}
\section{Where are high velocity clouds?}
\vspace{-5pt}

Wakker \& van Woerden (1997) have expounded on the complex history 
associated with high-velocity clouds. It is difficult to assign their true 
importance to astrophysics without a mean distance to the population. For 
example, a mean distance of (5, 50, 500) kiloparsecs leads to a total 
\HI\ mass of roughly ($10^7$, $10^9$, $10^{11}$) $M_\odot$.

Bland-Hawthorn \& Maloney (2001) show that 
the H$\alpha$ emission measures of HVCs
[E$_m$(obs)] are broadly consistent with the BM99$a$ model
[E$_m$(model)]. But the model does {\it not} explain the Magellanic
Stream detections, as we discuss in the next section. Our analysis
includes all HVCs with at least one known distance bound, as
summarized by Wakker (2000), with the exception of 5 HVCs which are
within 10\deg\ of the Galactic plane. The emission measures are from
the Las Campanas (Weiner et al.\ 2000), WHAM (Tufte et al.\ 1998;
Haffner 2000) and TAURUS surveys (Putman et al.\ 2000).
To within a factor of a few, an escape fraction of $f_{\rm esc} =$ 6\%
is consistent with the observed emission measures.  The predicted
emission measures arise from calculating the expected flux over
different cloud facets which see the disk, for a range of distances
within the allowed constraints. (For clouds with upper distance
bounds, we have adopted a lower bound of 0.5 kpc.)  We emphasize that
the predictions are at best broadly descriptive since the Galaxy is
modelled with a smooth exponential disk. There must be large local
variations in H$\alpha$ due to line-of-sight effects, limb brightening,
unrelated structures at low latitude, and so on.
For our crude estimate of $\hat{f}_{\rm esc}$ to be valid, it is crucial
that future H$\alpha$ surveys of nearby HVCs show the influence of the
Solar Circle (\S 5). This is a necessary but not a sufficient
condition. We would also expect to see the influence of the spiral
arms along their tangent points in longitude ({\it cf.} Fig.~2$a$). If
the Solar Circle is evident in the data, but not the spiral arms, this
might argue that $\hat{f}_{\rm esc}$ is much less than 6\%, and that
something like the EUV field of Slavin et al.\ (2000) dominates the
ionization.

If the H$\alpha$ normalization to local HVCs is valid, this may indicate
that some HVCs which are faint or undetected in H$\alpha$ (Weiner et al.\ 2000; 
Putman et al.\ 2000), particularly those at high latitude, are probably 
dispersed throughout the extended halo on scales of 50~kpc or more.

\vspace{-10pt}
\section{The Magellanic Stream.}
\vspace{-5pt}

\noindent{\bf Distance problem.}
Do we {\it really} know the distance and overall distribution of
Stream \HI\ in the absence of stellar probes along or close to the
Stream?  Who is to say that, seen from a vantage point 3~Mpc distant,
the Galaxy's environs do not resemble something like the complex
\HI\ network in the M81/M82 group (Yun, Ho \& Lo 1994). 

Our `intuition' is strongly guided by a slew of dynamical models,
which show tidal tails extending from either the SMC\footnote{The
canonical distances for the LMC and SMC are 50 and 60 kpc respectively
although Udalski et al.\ (1998) find that both are probably 15\%
closer.}  or from the LMC-SMC Lagrangian point (Gardiner \& Noguchi
1996; Li 1999). In other words, we are strongly dependent on intuition
guided by numerical models (Gibson et al.\ 2000).  However, few of
these models properly treat the gas or otherwise we would understand
why the Magellanic Stream appears to have no stellar counterpart (de
Vaucouleurs 1954; Recillas-Cruz 1982; Tanaka
\& Hamajima 1982; Br\"{u}ck \& Hawkins 1983; Westerlund 1990), although
it is possible to contrive models which separate gas and stars along 
tidal arms (Yoshizawa \& Noguchi 1999). What we can say is that the LMC-SMC
binary interaction within the extended dark halo of the Galaxy is highly
complex and we are far from a detailed understanding of it.

\begin{figure}
(a)
\centerline{ \rotatebox{-90}{\includegraphics[width=5.5cm]{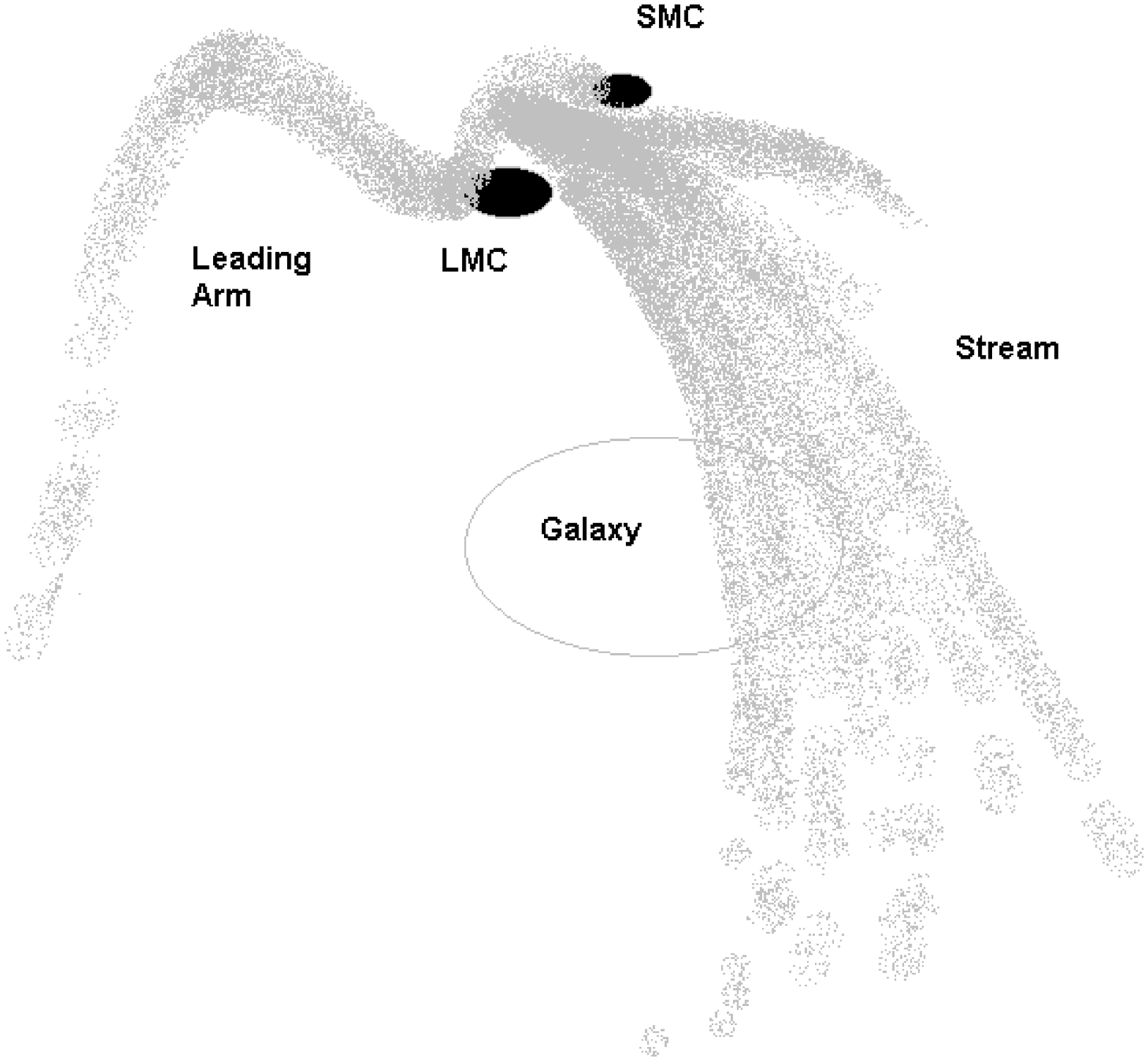}} }
(b)
\centerline{ \rotatebox{-90}{\includegraphics[width=5.5cm]{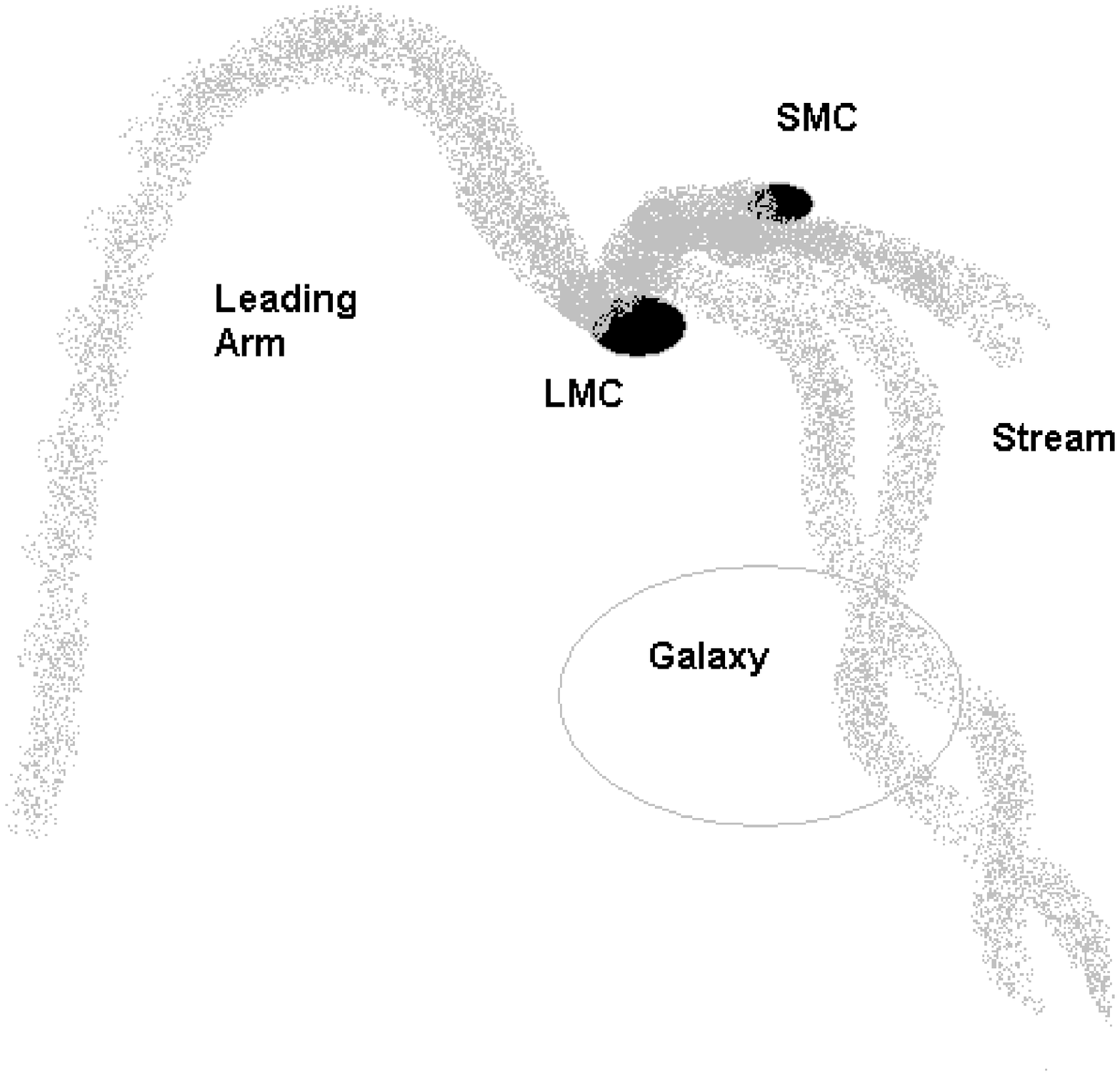}} }
(c)
\centerline{ \rotatebox{-90}{\includegraphics[width=5.5cm]{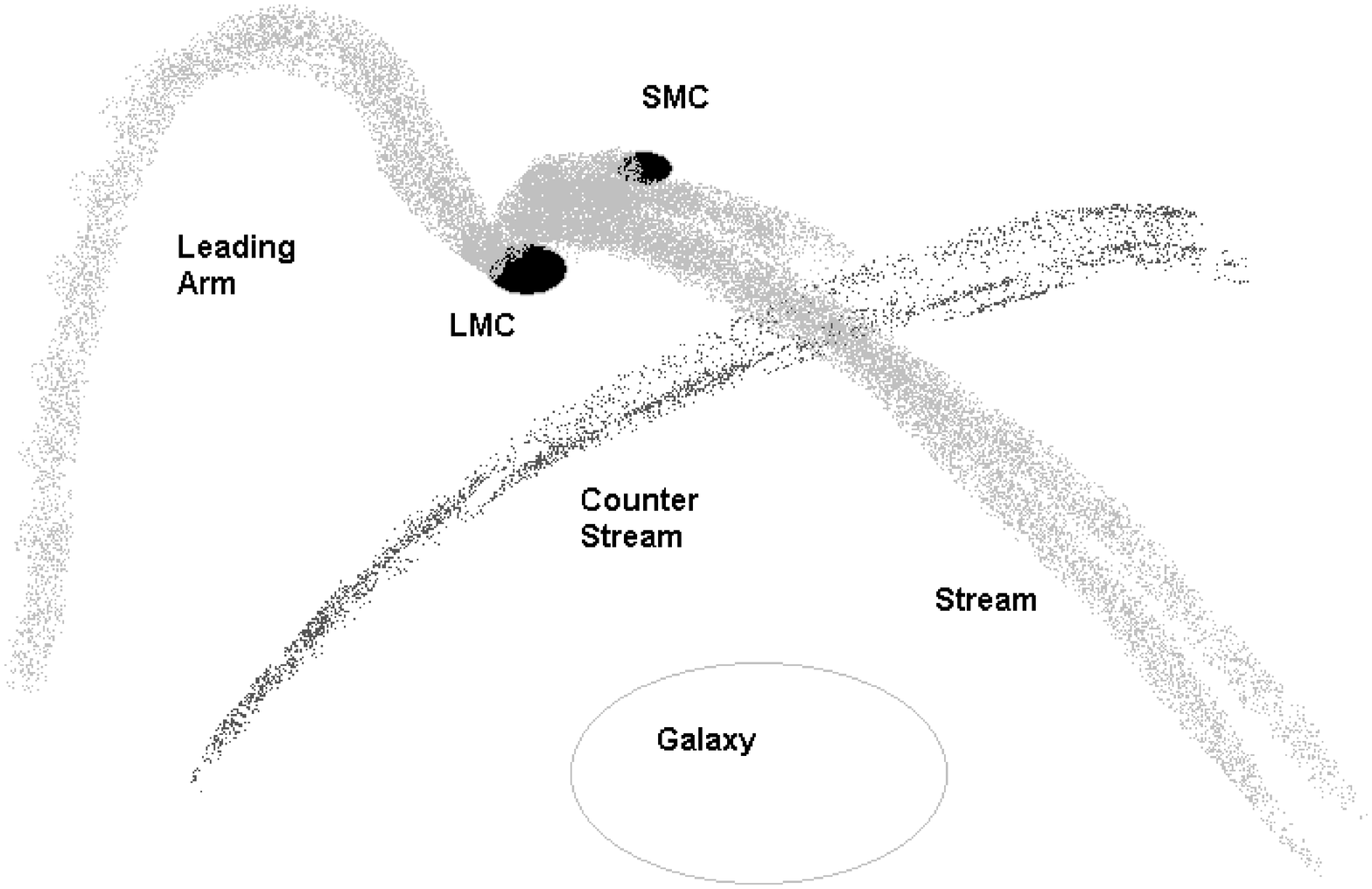}} }
\caption{\small Three possible configurations for the Magellanic Stream 
in order to explain the observed \HI/H$\alpha$ connection: 
(a) dispersed along the line of sight, (b) braided, (c) crossed.}
\label{bh2_fig3}
\end{figure}

Better distance estimates to different parts of the Stream may come from 
halo RR Lyrae or blue horizontal branch stars using the foreground-background 
technique to establish distance brackets. These can be easily picked 
out to at least 100~kpc with high quality photometry.

In Fig.~\ref{bh2_fig3}, we show three plausible configurations for the Stream,
each of which illustrates a point. Fig.~\ref{bh2_fig3}$a$ presents a Stream 
that is highly dispersed in a plane where the inner edge of the Stream
is much closer (say 20 kpc) than the assumed mean distance of 55 
kpc.\footnote{Note that models involving viscosity predict that the 
`tip' of the Stream  at MS~VI can be at least as close as 20 kpc 
(Moore \& Davis 1994).}  This configuration may help to explain 
two observations: (i) the exceptionally strong H$\alpha$ emission along
certain sight lines (\S 7), and (ii) the appearance of \HI\ lanes in 
Fig.~\ref{bh2_fig1} arising from tangent points due to slight undulations 
or structure within the Stream. The advantages of the other configurations 
are discussed below.

\smallskip
\noindent{\bf Ionization problem.}
A nagging problem with H$\alpha$ distances is the failure to explain the 
brightest H$\alpha$ detections along the Stream. For example, Weiner et al.\ 
(2000) have detected emission measures at MS~II as high as 400~mR, 
compared with 25--40~mR from the B99$a$ model at that same position.
Even the most contrived models may fail to patch up this discrepancy
since the cloud would need to be $3-4$ times closer (in the absence
of limb brightening).

Since the stellar searches to date have been limited in scope, 
young massive stars may have been missed along 
the Stream and these could explain some of the H$\alpha$ emission.
A case in point is the Shapley stellar wing in the Magellanic Bridge
(Courtes et al.\ 1995, Fig.~1) which has been shown to include stars
with ages less than 16~Myr ({\it q.v. } Demers \& Battinelli 1998; 
Rolleston et al.\ 1999). This region is very bright in H$\alpha$ 
($\sim$3R; Marcelin, Boulesteix \& Georgelin 1985; Veilleux et al.\ 2000). 

A few of the \HI\ clouds in the Stream appear to have head-tail 
morphologies.  Originally, Weiner \& Williams (1996) suggested possible H$\alpha$ limb 
brightening ahead of the Stream clouds giving support for shocks,
but new data makes this proposition less likely 
(Weiner et al.\ 2000). In fact, almost any overdensity contrast 
with the surrounding gas will confine the
cloud, i.e.\ a far wider parameter space than the narrow range of parameters
which produces optical shock emission (Murali 2000). 

The radiative regions in shocks are in pressure equilibrium 
with the external gas (Sutherland \& Dopita 1993) such that
$n_{\scriptscriptstyle\rm A} v_{\scriptscriptstyle\rm LMC}^2
\approx n_{\scriptscriptstyle\rm S} v_{\scriptscriptstyle\rm S}^2$
where $n_A$ is the ambient density, $v_{\scriptscriptstyle\rm LMC}$ is
the speed of the Stream in the frame of the Galaxy,
$v_{\scriptscriptstyle\rm S}$ and $n_{\scriptscriptstyle\rm S}$ are
the shock velocity and the post-shock density. We adopt a coronal
density of $n_{\scriptscriptstyle\rm A} \approx 10^{-4}$, the maximum
allowed by pulsar dispersion measures (B99$a$); at the head of clouds
MS II--IV, the volume-averaged atomic density from the HIPASS
observations is in the range $n_{\scriptscriptstyle\rm S} = 0.1-1$
cm$^{-3}$.  The Stream emission measures produce electron
densities in our range for any reasonable path length. Proper motion
studies indicate that the total Galactocentric transverse velocity for
the LMC is $v_{\scriptscriptstyle\rm LMC} = 213\pm 49$ km s$^{-1}$ (Lin,
Jones \& Klemola 1995).  The predicted shock velocities arising from
the Stream dynamics are only a few km s$^{-1}$, which are not enough to
ionize hydrogen.

So what about self-interaction? Cloud collisions of 20 km s$^{-1}$ or more 
produce H$\alpha$ through collisional {\it excitation} (with partially 
suppressed H$\beta$ emission relative to H$\alpha$).  If we could arrange 
to bang together \HI\ clouds at 80 km s$^{-1}$ or more,
collisional {\it ionization} makes life more interesting,
particularly if we allow for moderate levels of pre-ionization
by the Galactic disk. Fig.~\ref{bh2_fig3}$b$ and \ref{bh2_fig3}$c$ show two
interesting configurations.  Fig.~\ref{bh2_fig3}$b$ is a braided
trailing Stream arising from the binary orbit of the LMC--SMC
system. Fig.~\ref{bh2_fig3}$c$ has the Stream colliding with either
its own tail (since the LMC-SMC system {\it must} precess within the
extended halo) or with the \HI\ stream of some other infall object
({\it cf.} Putman et al.\ 2000). This picture is appealing because CDM
advocates propose that the Galactic halo is made up of hundreds or
even thousands of debris streams from accretion of small stellar
systems (Wyse 1999), some of which were presumably gas rich.

What are the likely shock signatures?  The post-shock velocity should
lead to a detectable offset from the \HI\ cloud. If the MS clouds
really are limb-brightened at the head of the shock, then this offset
might not be detectable since the clouds are overhead, but the
bowshock curvature should produce a detectable asymmetric wing in the
line profile (20 km s$^{-1}$ resolution or better).

If the shock velocity is less than 100 km s$^{-1}$, then ionization is 
produced in the shock itself, and there is a large collisional 
contribution to the Balmer lines and the 2-photon continuum
(Sutherland \& Dopita 1993; Shull \& McKee 1979).  Every ionized atom 
will make one recombination going through the shock, and  this
provides the H$\alpha$ flux.  (At somewhat higher velocity, collisional 
excitations of the neutrals become important.)

In order of increasing shock velocity ($20\rightarrow 100$ km s$^{-1}$),
the progression in well known optical diagnostics is:
\begin{verse}
large Balmer decrement, strong [\oi];\\
large Balmer decrement, strong [\oi] and [\sii];\\
normal Balmer decrement, [\nii] and [\oii] becoming strong;\\
normal Balmer decrement, [\oiii] appears.
\end{verse}
When interpreting the conventional shock diagnostics,
one must keep in mind that the gas-phase $\alpha$-elements  
(as judged from SMC) are $4-5$ times lower than Solar abundance, and
secondary products (e.g.\ N) are suppressed compared to the
$\alpha$-elements by a similar factor (Gibson et al.\ 2000). This calls for 
long exposures on the low ionization lines to be sure of reaching the 
necessary sensitivity. 

{\sl We note that partial pre-ionization by the Galactic disk can
assist the shock process since it lowers the required shock velocity
to achieve a given post shock temperature } ({\it cf.} Shull \& McKee
1979). Could differential cloud-cloud motions along the Stream be
sufficient to generate local H$\alpha$ enhancements?  The
configuration in Fig.~\ref{bh2_fig3}$b$ might be able to produce
(adiabatic) shock velocities of 20 km s$^{-1}$. However, within the
errors, the \HI\ and H$\alpha$ projected velocities appear to be the
same.

In fast shocks, the shock-generated UV spectrum can ionize the gas
ahead of the shock.  A shock velocity of 100 km s$^{-1}$ is needed to produce
nearly complete pre-ionization from the shock itself. At even higher
shock velocities ($>$175 km s$^{-1}$), an equilibrium \HII\ region is produced
ahead of the shock. Here, we would need a configuration like
Fig.~\ref{bh2_fig3}$c$ where \HI\ debris trails are on very different
trajectories.

\smallskip
\noindent{\bf Simple test.} The H$\alpha$ distribution along the 
Stream could provide the fundamental clue. If H$\alpha$ peaks at the poles,
this indicates knowledge of the Galaxy either through the presence of
disk photoionization (B99$a$), shock pre-ionization, a Galactic
wind/fountain, or whatever. If H$\alpha$ is bright at large angles from the
poles (e.g. MS~V--VI), this argues for something like
Fig.~\ref{bh2_fig3}$b$ and against a dominant Galactic influence,
unless the tip happens to be much closer to the disk.  Any strong H$\alpha$
asymmetry in Galactic coordinates argues for shock interactions
similar to Fig. ~\ref{bh2_fig3}$c$.

\vspace{-10pt}
\section{Future studies}
\vspace{-5pt}

We began this overview by acknowledging that the IGM has become one of
the main frontiers of modern astrophysics, largely driven by exquisite
absorption-line data from quasar spectra. This is a difficult topic
which will require careful study over many years.  Progress will come
from tackling the problem from many directions, starting with our own
backyard, i.e.\ the Galactic halo and the Local Group medium.

A full understanding of the halo and the role of HVCs will be slow in
coming since the physical processes are undoubtedly complicated
(e.g. Wolfire et al.\ 1995).  Hot gas has been detected along many
sight lines through the halo ({\it q.v. } Sembach et al.\ 1998). This
presumably arises from cooling hot gas becoming opaque to its own
radiation field, thereby indicating a complex multi-phase halo.  A
rather exotic possibility is halo material interacting with the IGM as
the Galaxy sweeps through the Local Group.  Important clues will come
from observing external galaxies much like our own. For example, in
NGC 5755, while there is evidence for a large-scale, smoothly
distributed source of halo ionization, the amplitude and variation in
[\oiii]/[\nii] clearly indicates secondary sources (Collins \& Rand
2001).

We can anticipate help from unexpected sources.  The Square Kilometre Array 
should detect pulsars in Local Group galaxies and thus provide plasma 
densities within the warm intergalactic medium (MB). The Sloan Digital Sky 
Survey should be able to identify stellar probes in the outer halo  on
100 kpc scales, and provide a distance ladder of stellar probes at
intermediate distances.  Future space astrometry missions may reveal
debris trails from hundreds of disrupting stellar satellites (`spaghetti
halo'), and we can foresee that the orbital parameters of these may account
for some of the HVC population.

We suspect that a reliable determination of $f_{\rm esc}$ for the
Galaxy is a necessary first step in understanding the halo ISM. An
interesting side product of $f_{\rm esc}$ (assuming greater than 1\%
or so) is a crude distance constraint to \HI\ clouds through the H$\alpha$
emission, even if only to clearly indicate which clouds are or are not
within the Galactic sphere of influence.  Fully convincing models of
the Stream interaction may require essentially complete H$\alpha$ maps
along its length at similar resolution to the HIPASS \HI\ maps. But
future absorption line studies using background probes will be crucial
for revealing how much of the story is taking place at low electron
and neutral columns ($< 10^{16}$ particles cm$^{-2}$), i.e.\ whether
the Stream is largely confined to the famous \HI\ arc, or whether it
extends over a much greater solid angle ({\it cf.} Gibson et al.\
2000).

\acknowledgments
We are indebted to Brad Gibson for his insights and help with all aspects
of this work.  JBH thanks Chris McKee for his combat and continued 
inspiration in this field. We benefitted from dialogues with Phil 
Maloney, Sylvain Veilleux, Ben Weiner, Rich Rand, Mark Giroux and 
Jonathan Slavin.

\bigskip
{\parindent 0pt

{\bf DISCUSSION:}

\medskip

\medskip{\bf Ron Allen:}
It is intriguing how often the
most energetic \HII\ regions generally seem to be `naked' (e.g.
M81; Allen et al.\ 1997). But we've known this since the old WSRT
maps of thermal emission in nearby galaxies showed no bright
discrete sources which were not also visible as \HII\ regions in H$\alpha$.

\medskip{\bf Todd Tripp:}
Nice evidence for UV flux escaping from galaxies (at high redshift)
has been provided by the He$^+$ Gunn-Peterson observations presented by
Alain Smette. He showed regions in the spectra of high redshift QSOs
in which the \HI\ opacity is extremely low and yet the \heii\ opacity is
extremely high. This suggests that these regions are ionized by very
soft sources, i.e.\ not quasars. UV flux escaping from, say, a
star-forming galaxy provides an appealing explanation. 

\smallskip{\bf Joss Bland-Hawthorn:} There is some support for Smette's 
argument from recent observations of Lyman-break galaxies: these
appear to be a significant source of UV flux (see Steidel et al.\
2000).

\medskip{\bf Sergei Marchenko:}
If most UV photons escape from \HII\ regions through chimneys, then the
external UV field should be very patchy which makes the H$\alpha$
measurements as a distance indicator for HVCs questionable. Is this
correct?

\smallskip{\bf Joss Bland-Hawthorn:} 
That depends on the opening angle of the chimneys.  If they really are
vertical tubes, then only a few HVCs are expected to light up, i.e.\
those caught in the searchlight beams. But if the chimneys have
reflecting walls (like skylights commonly used in Australian
households) or are slightly conic, then the halo field will become
uniform at some distance above the plane (related to the mean spacing
between the UV sources). Observations already show large opening
angles above the most powerful star-forming regions (Veilleux, this
meeting), but therein lies the rub.  These complexes are often widely
spaced over the disk.

}

\end{document}